\documentclass[a4paper,11pt]{article}
\usepackage{pos}
\usepackage{graphicx}
\usepackage{siunitx}
\usepackage{hyperref}
\usepackage{xspace}
\usepackage{lineno}

\newcommand{\reffig}[1]{Fig.~\ref{#1}}

\newcommand{\refsec}[1]{Section~\ref{#1}}

\newcommand{\refref}[1]{Ref.~\cite{#1}}

\newcommand{\Nmu}{\ensuremath{N_\mu}\xspace}
\newcommand{\Xmax}{\ensuremath{X_\mathrm{max}}\xspace}

\title{Plans for a new array of radio antennas for the detection of air showers at the 433\,m surface-detector array of the Pierre Auger Observatory}
\ShortTitle{A new array of radio antennas at the SD-433 of the Pierre Auger Observatory}

\author*[a]{Stef Verpoest}
\author[a]{Frank Schroeder}
\author[a]{Alexander Novikov}
\author[a]{Alan Coleman}
\author[a]{Benjamin Flaggs}
\author[b]{Andreas Weindl}
\author[b]{Megha Venugopal}
\author[b]{Carmen Merx}
\author[b]{Andreas Haungs}

\affiliation[a]{Bartol Research Institute, Department of Physics and Astronomy, University of Delaware,\\ Newark, DE 19716, USA}

\affiliation[b]{Institute for Astroparticle Physics (IAP), Karlsruhe Institute of Technology (KIT),\\76021 Karlsruhe, Germany}

\emailAdd{verpoest@udel.edu}
\emailAdd{fgs@udel.edu}

\abstract{We present the design and science case for a new array of radio antennas to be located at the Pierre Auger Observatory. Six stations of three SKALA antennas each will be deployed around a single water-Cherenkov surface detector triggering the radio readout. The planned antenna layout will allow for the detection of cosmic rays above a few tens of PeV and reach full efficiency for vertical air showers at several hundred PeV in primary energy. The array will thus be a pathfinder to demonstrate that fully-efficient radio detection in combination with the underground muon detectors already present at the location is possible. This will enable combined studies of the muon component and the depth of shower maximum, relevant for hadronic interaction models studies and more accurate determination of the cosmic-ray mass composition in the energy range of the Galactic-to-extragalactic transition.}

\FullConference{7th International Symposium on Ultra High Energy Cosmic Rays (UHECR2024)\\
 17-21 November 2024\\
Malargüe, Mendoza, Argentina\\}

\begin{document}
\maketitle

\section{Introduction}\label{sec:intro}

High-energy cosmic rays interacting in the Earth's atmosphere produce large particle cascades, known as extensive air showers.
The deflection of charged particles in the geomagnetic field and a time-varying negative charge excess at the shower front produce broadband radio emission, which can be observed with antennas at the ground.
Over the last two decades, radio detection has become a mature technique to study cosmic-ray air showers~\cite{Huege:2016veh,Schroder:2016hrv}. 
Complementary to ground-based particle detectors, radio antennas provide a direct measure of the cosmic-ray energy and the longitudinal shower development.

The Pierre Auger Observatory~\cite{PierreAuger:2015eyc} in Argentina is the largest cosmic-ray detector in the world, instrumenting over \SI{3000}{\km\squared} with over 1600 water-Cherenkov detectors, overlooked by fluorescence telescopes.
Recently, the installation of radio antennas on top of the water-Cherenkov detectors for the observation of horizontal air showers was finished as part of the AugerPrime upgrade, constituting the Auger Radio Detector (RD) array~\cite{PierreAuger:2023gql}.
Before that, the Auger Engineering Radio Array (AERA)~\cite{PierreAuger:2012ker} already demonstrated the maturity of the radio technique through e.g. a measurement of the depth of shower maximum ($X_\mathrm{max}$), finding results compatible with fluorescence measurements~\cite{PierreAuger:2023lkx,PierreAuger:2023rgk}. %Cite both papers
Radio upgrades for improved cosmic-ray detection are also ongoing at the IceCube Neutrino Observatory~\cite{IceCube:2016zyt}.
A planned enhancement consists of deploying scintillation detectors and radio antennas~\cite{Haungs:2019ylq} within the footprint of the surface detector, IceTop~\cite{IceCube:2012nn}.
Likewise, the future IceCube-Gen2~\cite{IceCube-Gen2:2020qha} project is planned to include a surface air-shower array consisting of the same detectors~\cite{IceCube-Gen2:2021aek, gen2surface}.

A prototype station of scintillator panels and radio antennas has been running at the South Pole for several years, demonstrating the detection of air showers in coincidence with IceTop~\cite{IceCube:2021epf,IceCube:2023rrl}.
Such a station was also deployed at the Pierre Auger Observatory and has successfully detected air showers in coincidence with the water-Cherenkov detectors of the Surface Detector (SD) array~\cite{Verpoest:2024fi}.
In the future, this will allow for cross checks between the observatories and may enable a cross calibration of the cosmic-ray energy scales.

In this contribution, we describe the plan to deploy more radio antennas at the Auger site, located within the SD-433, the densest part of the Auger SD array, with an inter-detector spacing of \SI{433}{\m}~\cite{PierreAuger:2021tmd}. The planned new array will consist of 18 antennas and may serve as a pathfinder for a larger array in the future.
The antennas will be deployed as a relatively dense array, complementing the sparser AERA and RD array by being sensitive mostly to vertical and lower-energy showers. It will also operate in a higher frequency band than the 30-\SI{80}{\mega\hertz} band used by AERA and the Auger RD.
The goal of this pathfinder array is to demonstrate the detection of near-vertical air showers with full efficiency, i.e. to be able to detect and successfully reconstruct all near-vertical showers above a certain energy threshold, given that they arrive within a certain distance from the array.
We also want to demonstrate the combined measurement of \Xmax and the muon content \Nmu in air showers, leveraging the Underground Muon Detector (UMD)~\cite{UMD:2023ZF} co-located with SD-433.
A combined measurement of \Xmax and \Nmu would provide improved mass-composition sensitivity~\cite{Holt:2019fnj,Flaggs:2023exc}, and would allow to perform tests of air-shower development and the hadronic interaction models used to simulate it.

\begin{figure}
    \centering
    \includegraphics[width=0.5\linewidth]{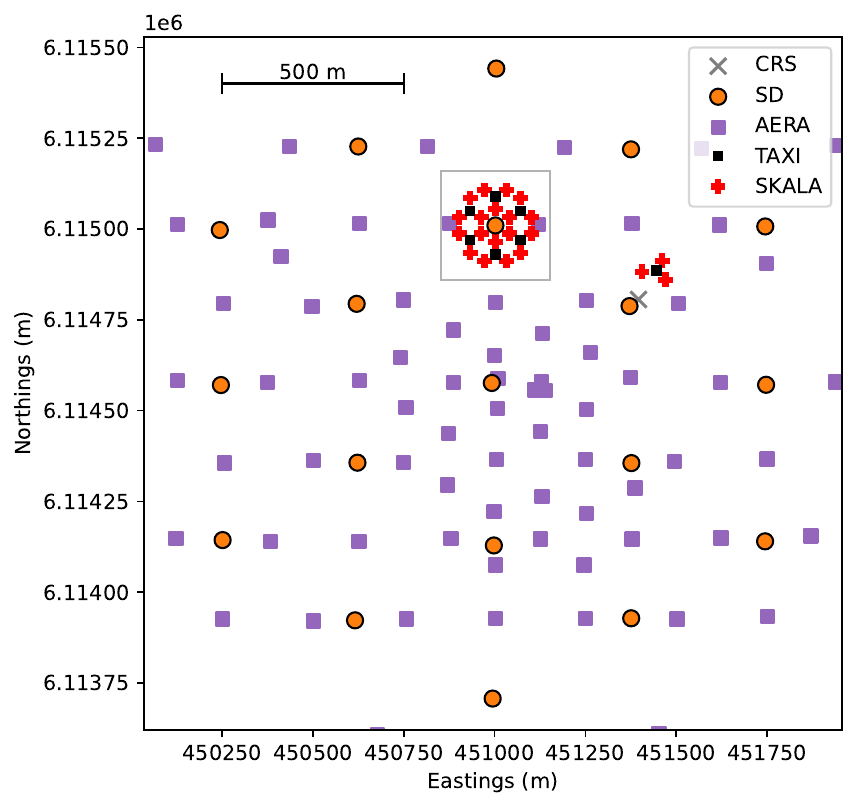}\includegraphics[width=0.5\linewidth]{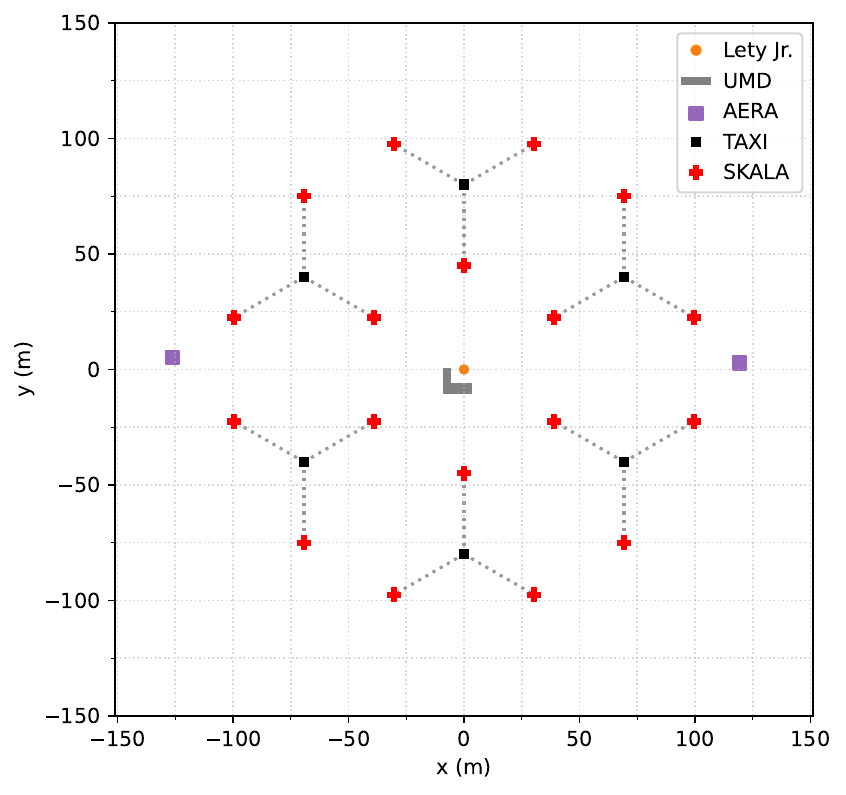}
    \caption{Planned locations for antenna deployment at the Pierre Auger Observatory. Left: Map showing the locations of the SD-433 detectors and AERA antennas. SKALA antennas and TAXI DAQs are shown by the red crosses and black squares respectively. The existing IceCube-Gen2 prototype station is shown near the Central Radio Station (CRS). The antennas inside the grey square are those planned to be deployed. Right: Zoom-in of the square containing the planned antenna locations from the left plot. The center of the plot represents the SD station, Lety Jr., around which the antennas will be deployed. Also shown are the nearby UMD modules. The dashed lines represent how the antennas are connected to the DAQ, defining which antennas belong to the same station.}
    \label{fig:array}
\end{figure}

The planned location and layout for the array are shown in \reffig{fig:array}. In the following sections, we discuss the technical design and optimization of the array layout in more detail.

\section{Array layout \& technical design}\label{sec:technical}

The planned layout for the array consists of 18 antennas grouped in six stations of three antennas each. 
The station design is inspired by the IceCube-Gen2 surface detector stations discussed in \refsec{sec:intro}.
The detector stations will consist out of a data-acquisition (DAQ) system, known as TAXI, to which three radio antennas are connected, with a nominal frequency band of \SI{70}{\mega\hertz} to \SI{350}{\mega\hertz} for the complete system.
The antennas are of the SKALA-v2 type and were originally designed for the SKA-low instrument~\cite{skala}.
In contrast to the prototype station for the IceCube surface array, there will be no scintillation detectors; the antenna readout will be triggered from the water-Cherenkov detector of a nearby SD station.
The six stations are planned to be deployed in a hexagon centered on a single SD station, \textit{Lety Jr.}, as shown in \reffig{fig:array}.
The exact layout, mainly the distance from Lety Jr. to the antenna-station centers and the distance of those station centers to the antennas, was optimized based on simulations, as discussed in \refsec{sec:simulation}.

\begin{figure}
    \centering
    \frame{\includegraphics[width=0.509\linewidth]{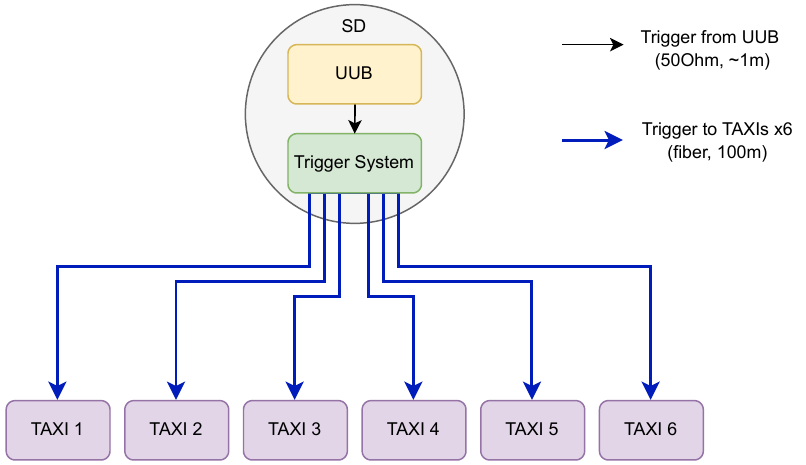}}\frame{\includegraphics[width=0.491\linewidth]{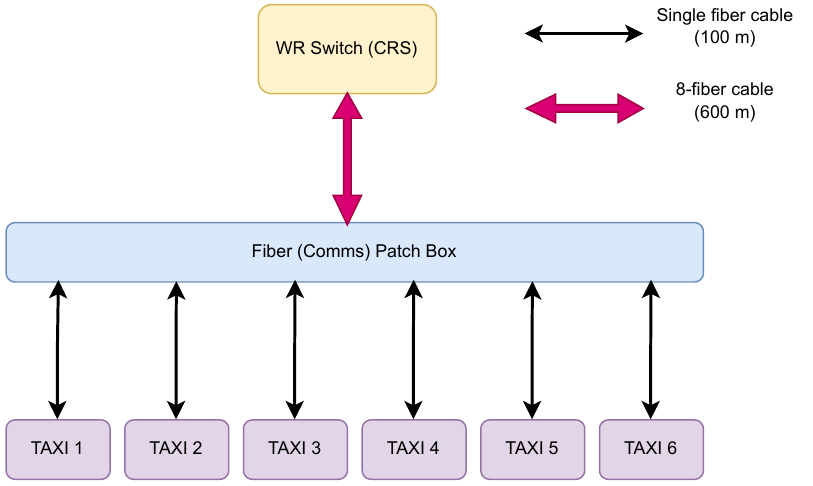}}
    \caption{Left: Schematic showing how the six TAXI DAQs are triggered simultaneously by the SD station using optical fiber. Right: Schematic showing how the TAXIs connect to a WhiteRabbit (WR) switch in the Central Radio Station (CRS) for timing and communication.}
    \label{fig:cabling}
\end{figure}

A local trigger from the central water-Cherenkov detector will be used for the simultaneous readout of all radio antennas.
The trigger signal is transferred from the Upgraded Unified Board (UUB) of the SD station, over six optical fibers, to each of the TAXIs, which have been modified to accept this external trigger (see \reffig{fig:cabling} (left)).
Filtering and merging of the events recorded on the individual TAXIs will happen offline by comparing to events with an SD array trigger, as a large fraction of the single-detector triggers will originate from events without any significant radio signal.

For timing and communication, the TAXIs will be connected to a WhiteRabbit switch~\cite{Lipinski2011WhiteRA}.
A single WhiteRabbit switch is located in the Central Radio Station of AERA.
An 8-fiber cable will run from the switch to a fiber patch box installed at the central SD station, where individual fibers will split off to connect to the TAXIs (see \reffig{fig:cabling} (right)).
The radio traces read out by the TAXIs will be transferred over these fibers and stored locally in the Central Radio Station for further offline filtering and processing. Each antenna station (TAXI + three antennas) will be powered autonomously using solar panels and batteries.

\section{Simulation study of the detection efficiency}\label{sec:simulation}

A simulation study was performed to optimize the exact layout of the hexagon of antenna stations.
The metric that was optimized for is the rate of near-vertical showers above the full-efficiency threshold for reconstruction. For simplicity, we considered only showers with cores contained within the antenna array. The problem therefore came down to optimizing the trade-off between a lower efficiency threshold and a larger array size.
Given the choice to deploy a hexagon of six stations around a single SD station, as described in \refsec{sec:technical}, only few degrees of freedom are left to optimize. These are the distance from the central SD station to the centers of the antenna stations, the distance from the antenna-station centers to the antennas, and the rotation of the antenna stations.

Air showers from proton primaries were simulated at fixed energies between $10^{16.5}\,$eV and $10^{18.25}\,$eV, and at fixed zenith angles from $0^\circ$ to $40^\circ$ in $10^\circ$ increments, while the azimuthal angle was chosen randomly.
CORSIKA~\cite{Heck:1998vt} and CoREAS~\cite{Huege:2013vt} were used to simulate the particle cascade and its radio emission.
We only simulated the response of the radio detector, as the full-efficiency threshold to trigger SD-433 of $10^{16.8}$\;eV is below the expected threshold for the radio reconstruction~\cite{PierreAuger:2023dju}.
We simulated the full response of the antenna and electronics chain, after which noise measured with the same antenna and electronics close to the deployment site was injected.
Each simulated air shower was resampled five times for the detector simulation, with a core position randomized within a circle centered on the central SD station and the radius defined by the furthest antenna.

Following the background frequency spectrum and air-shower analysis presented in \refref{Verpoest:2024fi}, simulated waveforms were bandpass filtered to \SI{110}{\mega\Hz} to \SI{185}{\mega\Hz}, after which a filter to suppress peaks in the background spectrum was applied~\cite{IceCube:2021qnf}.
To define the reconstruction efficiency, events containing signals with a high signal-to-noise ratio (SNR) in at least five antennas, as well as a successful directional reconstruction, were selected.
High SNR was in this case defined as the 95th percentile of the SNR distribution of measured background waveforms.
A plane-wave fit to the peak signal times in the antennas was then performed to obtain a simple reconstruction of the arrival direction.
The reconstruction is considered successful if the reconstructed direction has an opening angle of less than $5^\circ$ from the simulated direction.
We assume that for this selection of events, also an \Xmax reconstruction would be possible in the future.

The full efficiency threshold is defined as the energy above which 97\% of showers would be reconstructed, for each simulated zenith angle separately.
The total rate of events above full efficiency was calculated by integrating the efficiency over energy and zenith from the threshold upwards, assuming the H3a model for the cosmic-ray flux~\cite{Gaisser:2011klf}, and taking into account the solid angle and area of the circle in which showers were simulated.
The dependence of the efficiency on the azimuth was not treated explicitly.

The rate calculation was performed for SD-TAXI distances between \SI{60}{\m} and \SI{100}{\m}, TAXI-antenna distances between \SI{20}{\m} and \SI{40}{\m}, and rotations of $0^\circ$, $90^\circ$, and $180^\circ$, where $0^\circ$ corresponds to an antenna station having one TAXI-antenna arm pointing directly away from the central SD station.
As expected, denser layouts typically lowered the energy threshold, at the cost of a reduced detection area.
Several layouts with a similar performance were found. The final decision was made for an SD-TAXI distance of \SI{80}{\m}, a TAXI-antenna distance of \SI{35}{\m}, and a rotation of $180^\circ$, taking into account also practical deployment considerations. This is the layout shown in \reffig{fig:array}

\begin{figure}
    \centering
    \includegraphics[width=0.55\linewidth]{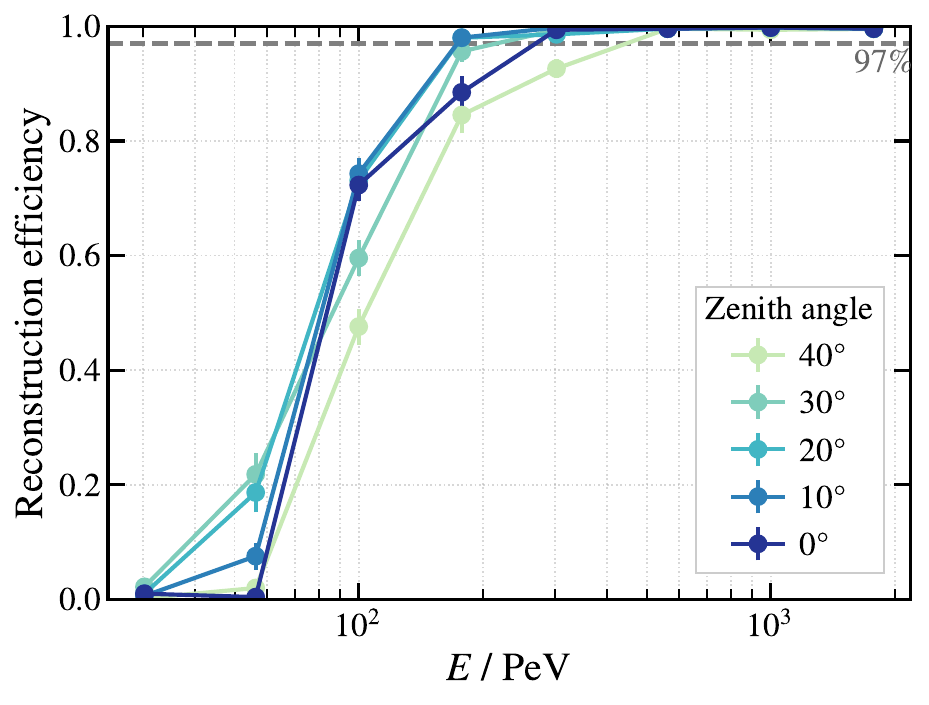}
    \caption{Expected reconstruction efficiency as a function of primary cosmic-ray energy for different simulated zenith angles. A shower is considered well-reconstructed if it has at least five antennas with a high-SNR signal and a directional reconstruction within 5$^\circ$ of the true value (see text for details). Only air showers with a core inside the circle defined by the outermost antennas of \reffig{fig:array} are considered.}
    \label{fig:efficiency}
\end{figure}

The reconstruction efficiency curves obtained for this layout are shown in \reffig{fig:efficiency}.
For showers up to 30 degrees in zenith, full efficiency is obtained around $10^{17.5}\,$eV or \SI{300}{\peta\eV}.
Around 45 events per year are expected above the full reconstruction efficiency threshold with the shower core contained within the antenna footprint, and a total number of about 500 events taking into account also the lower energy events.
Note that we only included near-vertical and contained events in these estimates; the total event rate is thus expected to be significantly higher.
We also note that the threshold based on the simple SNR cut can likely be lowered through improved filtering or the application of machine learning techniques, as in \refref{IceCube:2023rrl}.

\section{Conclusion and outlook}\label{sec:conclusion}

We have presented plans to deploy an array of SKALA antennas at the SD-433 array of the Pierre Auger Observatory.
The array will consist of 18 antennas, organized in six stations with a central DAQ each, deployed in a hexagon around a central SD station.
The readout of the antennas will be triggered by the water-Cherenkov detector of this SD station, and communication and timing will be provided through a connection to the AERA Central Radio Station.

We aim to demonstrate the radio detection of near-vertical air showers with full efficiency.
Simulations show that this is expected to be possible above cosmic-ray energies of several hundred PeV.
In addition, we aim to perform combined measurements of the depth of shower maximum and the muon number in the shower, by combining the radio measurements with measurements by the UMD.

The deployment of the antennas and DAQ hardware is expected to happen in early 2025.
Successful deployment and demonstration of the physics goals may encourage studies into further deployment of antennas at SD-433.
A larger dense array in this location could, through combined measurements of \Xmax and \Nmu, improve our understanding of cosmic-ray composition in the energy range of the Galactic-to-extragalactic transition, and allow for unique tests of hadronic interaction models.

\bibliographystyle{ICRC}
\bibliography{main}

\providecommand{\href}[2]{#2}\begingroup\raggedright\begin{thebibliography}{10}

\bibitem{Huege:2016veh}
T.~Huege \href{http://dx.doi.org/10.1016/j.physrep.2016.02.001}{{\em Phys. Rept.} {\bfseries 620} (2016) 1--52}.

\bibitem{Schroder:2016hrv}
F.~G. Schr\"oder \href{http://dx.doi.org/10.1016/j.ppnp.2016.12.002}{{\em Prog. Part. Nucl. Phys.} {\bfseries 93} (2017) 1--68}.

\bibitem{PierreAuger:2015eyc}
{\bfseries Pierre Auger} Collaboration, A.~Aab {\em et~al.} \href{http://dx.doi.org/10.1016/j.nima.2015.06.058}{{\em Nucl. Instrum. Meth. A} {\bfseries 798} (2015) 172--213}.

\bibitem{PierreAuger:2023gql}
{\bfseries Pierre Auger} Collaboration, J.~Pawlowsky {\em et~al.} \href{http://dx.doi.org/10.22323/1.444.0344}{{\em PoS} {\bfseries ICRC2023} (2023) 344}.

\bibitem{PierreAuger:2012ker}
{\bfseries Pierre Auger} Collaboration, P.~Abreu {\em et~al.} \href{http://dx.doi.org/10.1088/1748-0221/7/10/P10011}{{\em JINST} {\bfseries 7} (2012) P10011}.

\bibitem{PierreAuger:2023lkx}
{\bfseries Pierre Auger} Collaboration, A.~Abdul~Halim {\em et~al.} \href{http://dx.doi.org/10.1103/PhysRevLett.132.021001}{{\em Phys. Rev. Lett.} {\bfseries 132} no.~2, (2024) 021001}.

\bibitem{PierreAuger:2023rgk}
{\bfseries Pierre Auger} Collaboration, A.~Abdul~Halim {\em et~al.} \href{http://dx.doi.org/10.1103/PhysRevD.109.022002}{{\em Phys. Rev. D} {\bfseries 109} no.~2, (2024) 022002}.

\bibitem{IceCube:2016zyt}
{\bfseries IceCube} Collaboration, M.~G. Aartsen {\em et~al.} \href{http://dx.doi.org/10.1088/1748-0221/12/03/P03012}{{\em JINST} {\bfseries 12} no.~03, (2017) P03012}. [Erratum: JINST 19, E05001 (2024)].

\bibitem{Haungs:2019ylq}
{\bfseries IceCube} Collaboration, A.~Haungs \href{http://dx.doi.org/10.1051/epjconf/201921006009}{{\em EPJ Web Conf.} {\bfseries 210} (2019) 06009}.

\bibitem{IceCube:2012nn}
{\bfseries IceCube} Collaboration, R.~Abbasi {\em et~al.} \href{http://dx.doi.org/10.1016/j.nima.2012.10.067}{{\em Nucl. Instrum. Meth. A} {\bfseries 700} (2013) 188--220}.

\bibitem{IceCube-Gen2:2020qha}
{\bfseries IceCube-Gen2} Collaboration, M.~G. Aartsen {\em et~al.} \href{http://dx.doi.org/10.1088/1361-6471/abbd48}{{\em J. Phys. G} {\bfseries 48} no.~6, (2021) 060501}.

\bibitem{IceCube-Gen2:2021aek}
{\bfseries IceCube-Gen2} Collaboration, F.~Schroeder \href{http://dx.doi.org/10.22323/1.395.0407}{{\em PoS} {\bfseries ICRC2021} (2021) 407}.

\bibitem{gen2surface}
{\bfseries IceCube-Gen2} Collaboration, F.~Schroeder {\em PoS} {\bfseries UHECR2024} (these proceedings) 055.

\bibitem{IceCube:2021epf}
{\bfseries IceCube} Collaboration, H.~Dujmović {\em et~al.} \href{http://dx.doi.org/10.22323/1.395.0314}{{\em PoS} {\bfseries ICRC2021} (2021) 314}.

\bibitem{IceCube:2023rrl}
{\bfseries IceCube} Collaboration, A.~Rehman {\em et~al.} \href{http://dx.doi.org/10.22323/1.444.0291}{{\em PoS} {\bfseries ICRC2023} (2023) 291}.

\bibitem{Verpoest:2024fi}
{\textbf{IceCube-Gen2} and \textbf{Pierre Auger} Collaborations, S. Verpoest \textit{et al.}} \href{http://dx.doi.org/10.22323/1.470.0037}{{\em PoS} {\bfseries ARENA2024} (2024) 037}.

\bibitem{PierreAuger:2021tmd}
{\bfseries Pierre Auger} Collaboration, G.~Silli \href{http://dx.doi.org/10.22323/1.395.0224}{{\em PoS} {\bfseries ICRC2021} (2021) 224}.

\bibitem{UMD:2023ZF}
{\bfseries Pierre Auger} Collaboration, J.~de~Jesús \href{http://dx.doi.org/10.22323/1.444.0267}{{\em PoS} {\bfseries ICRC2023} (2023) 267}.

\bibitem{Holt:2019fnj}
E.~M. Holt, F.~G. Schr\"oder, and A.~Haungs \href{http://dx.doi.org/10.1140/epjc/s10052-019-6859-4}{{\em Eur. Phys. J. C} {\bfseries 79} no.~5, (2019) 371}.

\bibitem{Flaggs:2023exc}
B.~Flaggs, A.~Coleman, and F.~G. Schr\"oder \href{http://dx.doi.org/10.1103/PhysRevD.109.042002}{{\em Phys. Rev. D} {\bfseries 109} no.~4, (2024) 042002}.

\bibitem{skala}
E.~{de Lera Acedo}, N.~{Razavi-Ghods}, N.~{Troop}, N.~{Drought}, and A.~J. {Faulkner} \href{http://dx.doi.org/10.1007/s10686-015-9439-0}{{\em Experimental Astronomy} {\bfseries 39} no.~3, (Oct., 2015) 567--594}.

\bibitem{Lipinski2011WhiteRA}
M.~Lipinski, T.~Wlostowski, J.~Serrano, and P.~Alvarez {\em 2011 IEEE International Symposium on Precision Clock Synchronization for Measurement, Control and Communication} (2011) 25--30.

\bibitem{Heck:1998vt}
D.~Heck {\em et~al.} {\em Forschungszentrum Karlsruhe Report FZKA-6019} (1998) .

\bibitem{Huege:2013vt}
T.~Huege, M.~Ludwig, and C.~W. James \href{http://dx.doi.org/10.1063/1.4807534}{{\em AIP Conf. Proc.} {\bfseries 1535} no.~1, (2013) 128}.

\bibitem{PierreAuger:2023dju}
{\bfseries Pierre Auger} Collaboration, G.~Brichetto~Orquera \href{http://dx.doi.org/10.22323/1.444.0398}{{\em PoS} {\bfseries ICRC2023} (2023) 398}.

\bibitem{IceCube:2021qnf}
{\bfseries IceCube} Collaboration, A.~Coleman \href{http://dx.doi.org/10.22323/1.395.0317}{{\em PoS} {\bfseries ICRC2021} (2021) 317}.

\bibitem{Gaisser:2011klf}
T.~K. Gaisser \href{http://dx.doi.org/10.1016/j.astropartphys.2012.02.010}{{\em Astropart. Phys.} {\bfseries 35} (2012) 801--806}.

\end{thebibliography}\endgroup

\end{document}